# Multi-Feature Fusion-based Scene Classification Framework for HSR Images

Zhengrui Huang

*Abstract*—To realize high-accuracy classification of high spatial resolution (HSR) images, this letter proposes a new multi-feature fusion-based scene classification framework (MF²SCF) by fusing local, global, and color features of HSR images. Specifically, we first extract the local features with the help of image slicing and densely connected convolutional networks (DenseNet), where the outputs of dense blocks in the fine-tuned DenseNet-121 model are jointly averaged and concatenated to describe local features. Second, from the perspective of complex networks (CN), we model a HSR image as an undirected graph based on pixel distance, intensity, and gradient, and obtain a gray-scale image (GSI), a gradient of image (GoI), and three CN-based feature images to delineate global features. To make the global feature descriptor resist to the impact of rotation and illumination, we apply uniform local binary patterns (LBP) on GSI, GoI, and feature images, respectively, and generate the final global feature representation by concatenating spatial histograms. Third, the color features are determined based on the normalized HSV histogram, where HSV stands for hue, saturation, and value, respectively. Finally, three feature vectors are jointly concatenated for scene classification. Experiment results show that MF²SCF significantly improves the classification accuracy compared with state-of-the-art LBP-based methods and deep learning-based methods.

## I. Introduction

RECENT development in pattern recognition (PR) provides promising solutions to remote sensing (RS) [1], including object discovery and classification [2], land-cover change detection [3], biomass prediction and assessment [4], wetland restoration [5], etc, and the most basic part of RS is scene classification. Recently, high spatial resolution (HSR) images-based scene classification has attracted great attention, which can express the structure and texture of ground objects more clearly. To accurately classify HSR images, a large number of feature descriptors were proposed and developed [6], mainly including statistic-based methods (e.g., histogram of color (HoC) [7], gray level co-occurrence matrix (GLCM) [8], histogram of gradient (HoG) [9], local binary patterns (LBP) [10], and scale-invariant feature transform (SIFT) [2]) and learning-based methods (e.g., convolutional neural networks (CNN), deep CNN [11], long-short term memory (LSTM) [12], and generative adversarial networks (GAN) [13]), etc.

Interestingly, each kind of feature extractor has its unique receptive-field, which shows different advantages in different scales. Specifically, the statistic-based descriptors are good at extracting global (texture) features. The most representative method was original LBP that was firstly proposed for rotation invariant texture classification [10]. Based on the idea of LBP, some LBP-based variants were proposed and developed. In [14], local derivative pattern (LDP) was proposed to capture high-order derivative variations compared with LBP. From LDP, Zhang *et al*. designed a quantization thresholds-based local energy pattern (LEP) to further reduce the impact of imaging conditions. Other variants included local directional number patterns (LDNP) [15], dual-cross patterns (DCP) [16], local bit-plane decoded patterns (LBDP) [17], dominant rotated LBP (DRLBP) [18], etc.

However, these global feature descriptors can not effectively clarify local information of HSR images (e.g., key points, lines, or curves). Thus, the learning-based methods were introduced. The most widely used extractor was CNN or deep CNN [19], which could capture the detailed shape or structure information of objects. One well-known CNN architecture was VGGNet-16 proposed in [20], which consisted of 13 convolutional layers (Conv) and 3 fully connected layers (FC). Another variant of VGGNet was VGGNet-19 that contained extra three Convs. To ease the training process of VGGNet, a residual learning-based framework was proposed, and the most popular one was called ResNet-50 [21]. Following [21], a network architecture named densely connected convolutional networks (DenseNet) was proposed to improve propagation and reusing of feature-maps, which included $N$ dense blocks with $L(L+1)/2$ connections in each block, where $L$ was the number of layers [22]. Moreover, Inception-V3, Xception [23], and NASNet [24] were promising in PR community.

Excepted for local and global features, other kinds of image features should be considered, such as color information [25]. In [26], Zhang *et al*. proposed a color-based detection and matching method by analyzing the color distribution in HSV color space. To select the optimal color-components for object recognition, the best set of color features were generated from various color spaces, including YIQ, YQCr, and YCbCr [27]. Therefore, based on the characteristics of mentioned feature descriptors, we need to comprehensively decipher HSR images from various scales and levels.

Motivated by the above, we study in this letter a new scene

Zhengrui Huang is with the Academy of Digital China (Fujian), Fuzhou University, Fuzhou 350108, China, and the Key Laboratory of Spatial Data Mining & Information Sharing of Ministry Education, Fuzhou University, Fuzhou 350108, China.

classification framework by fusing the local, global, and color features of HSR images, namely, multi-feature fusion-based scene classification framework (MF²SCF), as shown in Fig. 1. Compared with the previous works [2]-[19], we consider the practical case that every feature extractor has a discriminative receptive-field, and the fusion of multiple feature extractors can retain image information as much as possible. Specifically, we propose three feature extractors to describe the local, global, and color feature of HSR images, respectively:

• Local feature extractor (LFE): In LFE, we jointly extract and connect the outputs of multiple dense blocks derived from the fine-tuned DenseNet-121 model to describe the global features of HSR images, as shown in Fig. 2. Especially, the inputs consist of two parts: 1) original HSR images and 2) HSR image slices. Image slicing technology can detect more detailed information from a part-image level [28], which includes 5 masks (each mask has 4 scales): 1) square 2) triangle 3) circle 4) left diagonal cropping (LDC) and 5) right diagonal cropping (RDC). Moreover, the outputs of dense blocks are processed through a global average pooling layer (GAP).

• Global feature extractor (GFE): GFE mainly consists of two steps: 1) complex networks-based (CN) mapping and 2) LBP encoding, as shown in Fig. 3. In the first step, we abstract a HSR image as an undirected graph based on pixel information, and obtain a gray-scale image (GSI), a gradient of image (GoI), and three CN-based feature images. In the second step, to reduce the impact of rotation and illumination on HSR images, we apply uniform LBP (ULBP) on GSI, GoI, and three feature images. Finally, the global feature is obtained by concatenating five spatial histograms.

• Color feature extractor (CFE): To capture the color feature of HSR images, we first transform original RGB color space to HSV color space. Second, the color feature representation of HSR images are obtained by joint normalization and HoC concatenation, as shown in Fig. 4.

The remainder of this letter is organized as follows. In Section II, we propose three feature extractors to describe HSR images. The experimental results and analyses are shown in Section IV, and the conclusions are drawn in Section IV.

*Notations*: Scalars are denoted by italic letters, and vectors and matrices are denoted by bold-face lower-case and bold-face upper-case letters, respectively. For a real $x$, $\mathbb{I}(x)$ stands for the indicator function that equals 1 if $x > 0$, and 0 otherwise. For a set $\mathcal{X}$, $|\mathcal{X}|$ denotes its cardinality.

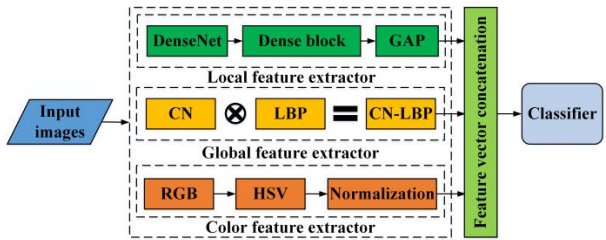

Fig. 1. Overview of MF²SCF.

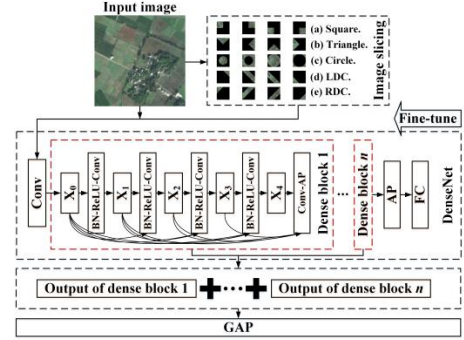

Fig. 2. Overview of LFE.

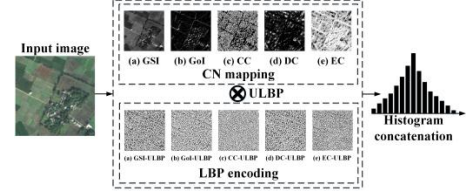

Fig. 3. Overview of GFE.

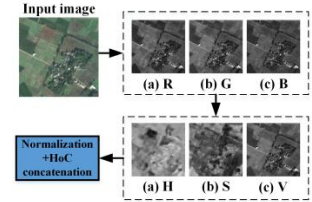

Fig. 4. Overview of CFE.

## II. METHODOLOGY

In this section, we introduce three feature extractors in order, including LFE, GFE, and CFE.

### A. LFE

Given a HSR image, denoted by $\mathbf{I}_0$, with size of $M \times M$, we first slice $\mathbf{I}_0$ into 20 slices (each mask has 4 scales, as shown in Fig. 2), denoted by $[\mathbf{S}_1,...,\mathbf{S}_{20}]$, which provides more detailed semantic information, and then input $[\mathbf{I}_0,\mathbf{S}_1,...,\mathbf{S}_{20}]$ into the fine-tuned DenseNet-121 model to extract local features.

Next, to capture the local features of $\mathbf{I}_0$ as much as possible, we fuse the outputs of dense blocks from DenseNet-121, and the input of layer $l$ in a dense block is given by [22]:

$$\mathbf{X}_l = \mathbb{H}([\mathbf{X}_0,...,\mathbf{X}_{l-1}]), l \in \mathcal{L} = \{0,...,L\} \qquad (1)$$

where $\mathbb{H}(\cdot)$ denotes the transformation operation, including a batch normalization layer (BN), a rectified linear unite layer (ReLU), and a Conv, $L$ is the number of layers, and the output of dense block $n$ can be expressed as:

$$\mathbf{Y}_n = \mathbb{T}(\mathbf{X}_L), n \in \mathcal{N} = \{0,...,N\} \qquad (2)$$

where $\mathbb{T}(\cdot)$ denotes the transition operation, including a average pooling layer (AP) and a Conv, and $N$ is the number of dense blocks.

Finally, based on (1)-(2), we concatenate multiple outputs of dense blocks in order, denoted by $[\mathbf{Y}_0,...,\mathbf{Y}_n]$, and express the final local feature vector as follows:

$$\mathbf{f}_1 = [\mathbb{G}(\mathbf{Y}_0),...,\mathbb{G}(\mathbf{Y}_n)], n \in \mathcal{N} \quad (3)$$

where $\mathbb{G}(\cdot)$ denotes the GAP, $|\mathbf{f}_1| = 2^6 + ... + 2^{6+n-1} + C$, and $C$ is the number of classes.

In this letter, to grab lower and higher activation values of feature maps in the DenseNet-121 model, we use a GAP to generate $\mathbf{f}_1$ instead of a FC. Consequently, LFE extracts local features from multi-scales (original HSR images and sliced HSR images) and multi-levels (multiple dense blocks), which can significantly strengthen the feature representation.

*B. GFE*

GFE mainly consists of two steps: 1) CN mapping and 2) LBP encoding. In the first step, we model $\mathbf{I}_0$ as an undirected graph by pixel information, including location, intensity, and gradient, and express the relationship (edge) between pixel $i$ and pixel $j$ (vertex), denoted by $e_{i,j}$, as follows:

$$e_{i,j} = \begin{cases} 1 & \text{if } w_{i,j} \leq t \text{ and } |g_i - g_j| \leq s \\ 0 & \text{otherwise} \end{cases}, i,j \in \mathcal{M} = \{0,...,M^2-1\} \quad (4)$$

where $w_{i,j}$ is the weight that equals $\left(\dfrac{d_{i,j}^2 + r^2 |I_i - I_j|/255}{2r^2}\right)$ if $d_{i,j} \leq r$, and 0 otherwise [29], $d_{i,j}$ is the pixel distance, $I$ is the intensity (gray-scale) value, $g$ is the gradient value that can be obtained by Sobel operator, and $r$, $t$, and $s$ are the thresholds of radius, similarity, and gradient difference, respectively.

Compared with traditional CN mapping in [29], we introduce the gradient-related constraint, and obtain the adjacent matrix of $\mathbf{I}_0$, denoted by $\mathbf{L}$. It is notable that $\mathbf{L}$ is a symmetric matrix, so we just need to traverse the pixels in the upper triangular matrix of $\mathbf{L}$, which reduces the computing complexity from $\mathcal{O}(M^4)$ to $\mathcal{O}(M^2 + (M^2-1) + ... + 1)$.

Since each HSR image has an unique adjacent matrix $\mathbf{L}$, we can discriminate two HSR images by comparing differences between their adjacent matrices. Here, three CN-based feature measurements are selected, including *clustering coefficient* (CC), *degree centrality* (DC), and *eigenvector centrality* (EC) [30], and used to express the spatial relationship, energy, and entropy of pixel $i$, respectively.

Specifically, the spatial relationship is given by CC:

$$\mathbb{C}(i) = \frac{2c_i}{k_i(k_i-1)} \quad (5)$$

where $k_i = \sum_{j=0}^{M^2-1} e_{i,j}$ denotes the degree of pixel $i$, and $c_i$ is the number of edges among neighbour pixels of pixel $i$. However, pixel energy and entropy can not be expressed by DC and EC directly [31], so we propose the following variants of DC and EC to express pixel energy and entropy, respectively:

$$\mathbb{D}(i) = \left(\frac{k_i}{M^2-1}\right)^2 \quad (6)$$

$$\mathbb{E}(i) = -\left(\lambda \sum_{j=0}^{M^2-1} e_{i,j} u_j\right)\left(\sum_{i=0}^{M^2-1} \log_2\left(\lambda \sum_{j=0}^{M^2-1} e_{i,j} u_j\right)\right) \quad (7)$$

where $\lambda$ is the reciprocal of the maximum eigenvalue of $\mathbf{L}$, and $u_j$ is the element of eigenvector associated with $\lambda^{-1}$.

Given (4)-(7), we can obtain a GSI, a GoI, and three feature images, which can describe the pixel intensity, gradient, spatial relationship, energy, and entropy, respectively. In the second step, to reduce the influence of rotation and illumination on HSR images, we apply ULBP on GSI, GoI, and three feature images to generate the global feature images:

$$\text{ULBP}(I_c) = \begin{cases} \sum_{p=0}^{P-1} \mathbb{I}(I_p - I_c) 2^p & \text{if } \mathbb{U}(\text{LBP}(I_c)) \leq 2 \\ P+1 & \text{otherwise} \end{cases} \quad (8)$$

where $I_c$ and $I_p$ are the gray-scale value of a central pixel $c$ and a neighbour pixel $p \in \mathcal{P} = \{0,...,P-1\}$, respectively, $P$ is the number of neighbour pixels, $\text{LBP}(I_c) = \sum_{p=0}^{P-1} \mathbb{I}(I_p - I_c) 2^p$, and $\mathbb{U}(\cdot)$ denotes the uniform pattern function [10]:

$$\mathbb{U}(\text{LBP}(I_c)) = |\mathbb{I}(I_{P-1} - I_c) - \mathbb{I}(I_0 - I_c)| + \sum_{p=1}^{P-1} |\mathbb{I}(I_p - I_c) - \mathbb{I}(I_{p-1} - I_c)| \quad (9)$$

Therefore, according to (4)-(9), we can obtain GSI-ULBP, GoI-ULBP, CC-ULBP, DC-ULBP, and EC-ULBP, as shown in Fig. 3, and five spatial histograms are jointly concatenated to form the final global feature vector:

$$\mathbf{f}_2 = [\mathcal{H}_{\text{GSI-ULBP}}, \mathcal{H}_{\text{GoI-ULBP}}, \mathcal{H}_{\text{CC-ULBP}}, \mathcal{H}_{\text{DC-ULBP}}, \mathcal{H}_{\text{EC-ULBP}}] \quad (10)$$

where $\mathcal{H}$ denotes the spatial histogram, and $|\mathbf{f}_2| = 59 \times 5 = 295$.

*C. CFE*

Except for the local and global feature, we also introduce the color feature to improve the discriminating ability of MF²SCF. In CFE, we do not use RGB channels to describe color space of HSR images. Instead, the HSV model is adopted for color feature extraction, which reflects how humans view color [32]:

$$H = \begin{cases} 0° & , \overline{C} = 0 \\ 60°\left(\dfrac{\overline{G}-\overline{B}}{\overline{C}} \% 6\right), C_{\max} = \overline{R} \\ 60°\left(\dfrac{\overline{B}-\overline{R}}{\overline{C}} + 2\right), C_{\max} = \overline{G} \\ 60°\left(\dfrac{\overline{R}-\overline{G}}{\overline{C}} + 4\right), C_{\max} = \overline{B} \end{cases}, S = \begin{cases} 0 & , C_{\max} = 0 \\ \dfrac{\overline{C}}{C_{\max}}, C_{\max} \neq 0 \end{cases}, V = C_{\max} \quad (11)$$

where $\overline{R} = R/255$, $\overline{G} = G/255$, $\overline{B} = B/255$, $\overline{C} = C_{\max} - C_{\min}$, $C_{\max} = \max\{\overline{R}, \overline{G}, \overline{B}\}$, and $C_{\min} = \min\{\overline{R}, \overline{G}, \overline{B}\}$.

To better the comparison of HSR images, we normalize (11) and express the final color feature vector as follows:

$$\mathbf{f}_3 = [\mathbb{N}(\mathcal{H}_H), \mathbb{N}(\mathcal{H}_S), \mathbb{N}(\mathcal{H}_V)] \quad (12)$$

where $\mathbb{N}(\cdot)$ is the minimum-maximum normalization function, and $|\mathbf{f}_3| = 256 \times 3 = 768$.

Given (3), (10), and (12), the feature vectors of HSR images are finally obtained, denoted by $[\mathbf{f}_1, \mathbf{f}_2, \mathbf{f}_3]$, and used for final scene classification.

### III. Experiments and Analyses

#### A. Datasets Description and Settings

In our experiments, we select three open source datasets to verify the effectiveness of MF²SCF, including SIRI-WHU [33], WHU-RS19 [34], and NWPU-RESISC45 [19]:

• SIRI-WHU: The dataset contains 12 classes, and each class has 200 samples (200×200 pixels).

• WHU-RS19: The dataset has 19 classes, and the number of samples (600×600 pixels) in each class ranges from 50 to 58.

• NWPU-RESISC45: The dataset has total 31500 samples divide into 45 classes, and the image size is 256×256 pixels.

In this letter, the parameters of GFE are given by: $P=8$, $r=3$, $t=0.315$, and $s=5$, and LFE is pretrained and fine-tuned for each dataset, i.e., we only adopt the backbone of the transferred DenseNet-121 model (not including its top classifier), and the number of epochs and the batchsize are equal to 10 and 32, respectively, and the model optimizer is root mean square prop (RMSProp), where the learning rate and the discounting factor equal 0.001 and 0.9, respectively. For final scene classification, the linear-kernel-based support vector machine (LK-SVM) is trained, where the penalty coefficient and the kernel coefficient are equal to 1.0 and the reciprocal of number of features, namely, $1/\|[\mathbf{f}_1, \mathbf{f}_2, \mathbf{f}_3]\|$, respectively.

Before training our model, we randomly split each dataset into three parts: 1) the first one is for fine-tuning DenseNet-121 model 2) the second one is for training LK-SVM and 3) the third one is for testing, where the training or testing samples are randomly selected, and the test size is equal to 0.3. Moreover, it is worth noting that all results are averaged over a large number of independent experiments through Python, including NUMPY, OPENCV, NETWORKX, KERAS, etc.

#### B. Comparison and Analyses of MF²SCF

In this section, the statistic-based descriptors and the deep CNN-based methods are compared with MF²SCF to evaluate the classification accuracies:

• Statistic-based descriptors: GLCM, HoG, LBP, LDP, LEP, LDNP, DCP, LBDP, DRLBP, and centre symmetric quadruple pattern (CSQP) [35].

• Deep CNN models: VGGNet-16/19, ResNet-50, DenseNet-121/169/201, Xception, and NASNet-Large.

To evaluate the performance of MF²SCF, the micro accuracy score is adopted to compute the differences between predicted and true labels. As shown in Table I, we present the results of classification accuracies, and can see that MF²SCF gives better performance, where the classification accuracies on three datasets reach 96.53%, 98.59%, and 90.22%, respectively. Compared with statistic-based descriptors and deep CNN models, our proposed MF²SCF retains more detailed semantic information by fusing the local, global, and color information, which better deciphers the characteristics of HSR images.

Moreover, MF²SCF has abilities to reduce the impact of illumination or imaging with the help of LBP encoding.

TABLE I
CLASSIFICATION ACCURACIES (%) OF MF²SCF, DIFFERENT STATISTIC-BASED DESCRIPTORS, AND DEEP CNN MODELS

|   | Method | SIRI-WHU | WHU-RS19 | NWPU-RESISC45 |
|---|---|---|---|---|
| ▲ | GLCM | 62.64 | 63.46 | 38.67 |
|   | HoG | 50.33 | 58.91 | 26.12 |
|   | LBP | 52.64 | 56.11 | 46.67 |
|   | LDP | 56.53 | 59.64 | 49.98 |
|   | LEP | 41.67 | 46.52 | 38.19 |
|   | LDNP | 91.63 | 92.52 | 83.17 |
|   | DCP | 87.26 | 88.04 | 79.26 |
|   | LBDP | 91.29 | 93.62 | 81.80 |
|   | DRLBP | 91.11 | 92.72 | 80.89 |
|   | CSQP | 89.85 | 90.28 | 78.38 |
| ▼ | VGGNet-16 | 91.11 | 95.09 | 73.78 |
|   | VGGNet-19 | 89.44 | 92.98 | 73.56 |
|   | ResNet-50 | 92.22 | 96.84 | 81.96 |
|   | DenseNet-121 | 93.52 | 97.19 | 83.26 |
|   | DenseNet-169 | 91.26 | 96.49 | 80.19 |
|   | DenseNet-201 | 90.31 | 95.78 | 78.86 |
|   | Xception | 76.05 | 79.64 | 58.66 |
|   | NASNet-Large | 83.68 | 83.88 | 61.23 |
|   | **MF²SCF** | **96.53** | **98.95** | **90.22** |

▲: Statistic-based descriptors ▼: Deep CNN models

Moreover, from the perspective of computing cost, since we jointly transfer and fine-tune the parameters of pre-trained DenseNets, MF²SCF has higher computing efficiency and can save more operation time compared with completely trained deep CNN-based models. As shown in Table II, we can find that the computing costs of MF²SCF on three datasets are 291 s, 1004 s, and 1342 s, respectively.

TABLE II
COMPUTING COSTS IN PER EPOCH (S) OF MF²SCF AND DEEP CNN MODELS

| Method | SIRI-WHU | WHU-RS19 | NWPU-RESISC45 |
|---|---|---|---|
| VGGNet-16 | 364 | 1285 | 1974 |
| VGGNet-19 | 423 | 1428 | 2087 |
| ResNet-50 | 341 | 1087 | 1536 |
| DenseNet-121 | 412 | 1397 | 1873 |
| DenseNet-169 | 486 | 1513 | 1967 |
| DenseNet-201 | 603 | 1744 | 2120 |
| Xception | 372 | 1340 | 1918 |
| NASNet-Large | 874 | 1898 | 3012 |
| **MF²SCF** | **291** | **1004** | **1342** |

To improve the robustness of MF²SCF, we further adopt dimension reducing (DC) methods to reduce the length of final feature vector, including principal components analysis (PCA) and linear discriminant analysis (LDA). In this letter, the feature numbers of PCA and LDA are determine by the curve of cumulative explained variance and the number of classes in each dataset, respectively. From Table III, we can know that PCA+MF$^2$SCF gives the best performance, where the optimal numbers of principal components of three datasets are equal to 420, 665, and 3047, respectively.

TABLE III
CLASSIFICATION ACCURACIES (%) OF MF$^2$SCF WITH DIFFERENT DC METHODS

| Method | SIRI-WHU | WHU-RS19 | NWPU-RESISC45 |
|---|---|---|---|
| MF$^2$SCF+PCA | 97.78 | 99.64 | 92.18 |
| MF$^2$SCF+LDA | 92.22 | 91.57 | 88.07 |

## IV. CONCLUSION

To realize accurate scene classification, this letter proposed a new scene classification framework named MF$^2$SCF by fusing local, global, and color features of HSR images. Specifically, we first extracted the local feature based on the transferred and fine-tuned DenseNets. Second, we jointly mapped and encoded GSI, GoI, and CN-based feature images to express the global feature. Third, we obtained the color feature of HSR images by HSV model-based normalization and HoC concatenation. The results demonstrated that MF$^2$SCF gave better classification accuracies on three datasets compared with the statistic-based methods and the deep CNN models.